\documentclass[preprint, superscriptaddress, showpacs,preprintnumbers,amsmath,amssymb,prb]{revtex4}
\usepackage{graphicx}

\begin{document}

\thispagestyle{empty}

\title{Impact of surface imperfections on the Casimir force for lenses of
centimeter-size curvature radii}

\author{V.~B.~Bezerra}
\affiliation{Department of Physics, Federal University of Para\'{\i}ba,
C.P.\ 5008, CEP 58059--900, Jo\~{a}o Pessoa, Pb-Brazil}

\author{G.~L.~Klimchitskaya}
%\email{Galina.Klimchitskaya@itp.uni-leipzig.de}
\affiliation{Department of Physics, Federal University of Para\'{\i}ba,
C.P.\ 5008, CEP 58059--900, Jo\~{a}o Pessoa, Pb-Brazil}
\affiliation{North-West Technical University, Millionnaya Street 5,
St.Petersburg, 191065, Russia}

\author{U.~Mohideen}
%\email{Umar.Mohideen@ucr.edu}
\affiliation{Department of Physics and
Astronomy, University of California, Riverside, California 92521,
USA}

\author{V.~M.~Mostepanenko}
%\email{Vladimir.Mostepanenko@itp.uni-leipzig.de}
\affiliation{Department of Physics, Federal University of Para\'{\i}ba,
C.P.\ 5008, CEP 58059--900, Jo\~{a}o Pessoa, Pb-Brazil}
\affiliation{Noncommercial Partnership ``Scientific Instruments'',
Tverskaya Street 11, Moscow, 103905, Russia}

\author{C.~Romero}
\affiliation{Department of Physics, Federal University of Para\'{\i}ba,
C.P.\ 5008, CEP 58059--900, Jo\~{a}o Pessoa, Pb-Brazil}

\begin{abstract}
The impact of imperfections, which are always present on surfaces
of lenses with centimeter-size curvature radii, on the Casimir force
in the lens-plate geometry is investigated. It is shown that the
commonly used formulation of the proximity force approximation is
inapplicable for spherical lenses with surface imperfections,
such as bubbles and pits. More general expressions for the
Casimir force are derived that take surface imperfections into
account. Using these expressions we show that surface imperfections
can both increase and decrease the magnitude of the Casimir force
up to a few tens of
percent when compared with the case of a perfectly spherical lens.
We demonstrate that the Casimir force between a perfectly spherical
lens and a plate described by the Drude model can be made approximately
equal to the force between a sphere with some surface imperfection
and a plate described by the plasma model, and vice versa.
In the case of a metallic sphere and semiconductor plate, approximately
the same Casimir forces are obtained for four different descriptions of
charge carriers in the semiconductor if appropriate surface imperfections
on the lens surface are present. The conclusion is made that there
is a fundamental problem in the interpretation of measurement
data for the Casimir force, obtained by using spherical lenses of
centimeter-size radii, and their comparison with theory.
\end{abstract}
\maketitle

\section{Introduction}

The  Casimir force \cite{1} is caused by the existence of
zero-point and thermal fluctuations of the
 quantized electromagnetic
field. In the last few years physical
phenomena grouped under the generic name of {\it Casimir
effect} have received much experimental and theoretical
attention \cite{2} owing to numerous prospective applications
in both fundamental and applied science.
Many theoretical results in Casimir physics (for instance,
on the role of skin depth or surface roughness)
have already been
experimentally confirmed (see Refs.\cite{17,19,9,26}
and review\cite{3}).
There is, however, one theoretical prediction made on the
basis of the Lifshitz theory \cite{2,4,5} which was
unexpectedly found to be in contradiction with the
experimental data. This is the large thermal effect in the Casimir
force at short separations caused by the relaxation properties
of free charge carriers in metals,\cite{6} semiconductors
and dielectrics.\cite{7,8}
The respective experiments performed by means of
micromechanical torsional oscillator,\cite{9,10,11}
atomic force microscope \cite{12} and Bose-Einstein
condensate confined in a magnetic trap \cite{13,14} excluded the
predicted effect at a high confidence level.

Almost all experiments on measuring the Casimir force between
two macroscopic bodies were performed in the sphere-plate
geometry.\cite{3}
Experiments exploiting the sphere-plate geometry can be
separated into experiments with small spheres of micrometer-size
radii \cite{17,19,9,26,10,11,12,18,20,21,22,23,24,25,27,28,29}
and with large spherical lenses of centimeter-size curvature
radii.\cite{30,31,29a,29aa}
In most of cases spherical surfaces were coated with Au
(single exception is the experiment\cite{31} using a lens made
of Ge).
Small spheres (from a few tens to hundreds of micrometer radii)
are usually made of polystyrene or sapphire.
It is possible to control both
global and local sphericity of small spheres microscopically by using, for
instance, scanning electron microscopy. Large spherical lenses
from a few centimeters to more than 10\,cm curvature radii
are made of glass or some other material.
Allowed parameters of imperfections (defects) of their
mechanically polished and ground surfaces are specified in the
optical surface specification data provided by a producer
(see, e.g., Refs.\cite{40,41,530a}).

It should be stressed that different defects are necessarily present
on the surface of each (even of the best quality) optical
lens.\cite{29b} The reason is that when the glass is first made
it may already contain defects such as bubbles. In grinding and
polishing, a whole new set of surface defects such as scratches, digs,
and chips may be introduced.\cite{29b} In the subsequent
technological operations of centering, beveling, cementing,
and assembly, more defects are likely to be produced.
The handling and operations involved in the numerous cleanings
and inspections also add their quota of defects.\cite{29b}
Because of this, such a specification as ``no bubbles or other
imperfections permitted'' is impossible to fulfil.\cite{29b}
Optical surface data specifying the parameters of
defects permissible are obtained using scanning scattering
microscopes, laser interference imaging profilometers and
other techniques.\cite{29c} The micrographs of different types
of defects of optical surfaces taken with a differential
interference contrast microscope can be found in
Ref.\cite{29d}. The scanning electron microscope images of
defects are contained in Ref.\cite{29e}.
The most frequently present imperfections on lenses are digs,
which include all hemispherical-appearing defects, and scratches
whose length is usually much longer than the wavelength
of the incident light.\cite{29b,29c,29d,29e,29f}
Note that in the large-scale applied problem of lens
design\cite{29g} surface inperfections play a rather
limited role. However, as is shown below, they are very
important for such a nonstandard application of lenses
as for measurements of the Casimir force.

It is important to bear in mind that although large thermal
corrections to the Casimir force at short separations were
experimentally excluded, the thermal effect by itself in the
configuration of two macrobodies has never been measured.
In this respect experiments with lenses of large curvature
radii attract much attention because they might allow
measurements at separations of a few micrometers where
predictions of alternative theoretical approaches
(taking into account or discarding relaxation properties of
free charge carriers) differ by up to 100\%.

Experiments on measuring the Casimir force using
spherical lenses of large
radius of
curvature have faced serious problems. The point is
that calibration of the Casimir setup is usually performed by
measuring electric forces between the sphere and plate from
a potential difference applied to the test bodies
(some nonzero residual potential difference exists even when the
test bodies are grounded). Calibrations are performed by the
comparison of the measured electric forces at different
separations with the exact theoretical force-distance relation
in the sphere-plate geometry, which is familiar from classical
electrodynamics. Problems emerged when an anomalous
force-distance relation for the electric force between an
Au-coated spherical lens of $R=3.09\pm 0.015\,$cm curvature
radius and a plate was observed,\cite{32} distinct from that
predicted by classical electrodynamics (see also Ref.\cite{33}).
The existence of anomalous electrostatic forces was also
confirmed in the configuration of Ge lens of
$R=15.10\,$cm curvature radius and Ge plate,\cite{31}
but denied \cite{34} for an Au-coated small sphere of
$R=100\,\mu$m radius interacting with an Au-coated plate.

It was shown \cite{35} that the anomalous behavior of the
electrostatic force can be explained due to deviations of the
mechanically polished and ground surface
 from a perfect spherical shape for lenses with
centimeter-size curvature radii.
Different kinds of imperfections on such surfaces (bubbles,
pits and scretches) can lead to significant deviations of the
force-distance relation from the form predicted by classical
electrodynamics under an assumption of perfectly spherical
surface. Later this possibility was recognized \cite{36} as
a crucial point to be taken into account in future experiments
not only in the sphere-plate geometry, but
also for a cylindrical lens
of centimeter-size radius of curvature near the plate.

In this paper we consider the possible imperfections on surfaces
of lenses with centimeter-size radius of curvature, and calculate
their impact on the Casimir force.
The point to note is that the Casimir force is far more sensitive
than the electrostatic force to the bubbles and pits that are
always present on the mechanically polished and ground
surfaces. The physical reason is that the Casimir force falls
with the increase of separation distance more rapidly than the
electric force. As a result, the Casimir force is determined
by smaller regions near the points of closest approach of the
two surfaces. If the local radius of curvature on the lens surface near
the point of closest approach to the plate is significantly
different from the mean radius of curvature $R$, the impact
of such surface imperfection on the Casimir force can be
tremendous.

We show that the presence of bubbles and pits on a lens surface,
allowed by the optical surface specification data,
 makes inapplicable the simplified
formulation of the proximity force approximation (PFA)
used \cite{30,31,29a} for the comparison between experiment and
theory. We also derive the expressions for the Casimir
force applicable in the presence of bubbles and pits on
surfaces of centimeter-size lenses. It is shown that for
ideal metal bodies surface imperfections may lead to both
a decrease and an increase in the magnitude of the Casimir
force up to a few tens of percent for
sphere-plate separations from 1 to $3\,\mu$m.

As discussed above, one might expect that experiments with
large lenses will help to resolve the problem with the
thermal Casimir force. In this connection we consider real
metal spherical lens, with surface imperfections of different
types, close to a real metal plate both described either by the
Drude dielectric function (relaxation of free charge carriers
is included) or by the dielectric function of the plasma
model where the relaxation parameter
of free charge carriers is set to zero.
We show, that the Casimir force between a perfectly
spherical lens and a plate, both described by the Drude
model, in the limit of experimental error, is equal to the
Casimir force between a lens with some specific surface
imperfection and a plate, both described by the plasma model.
Vice versa, we demonstrate that if the metal surface of the
perfectly shaped lens and a plate is described by the plasma
model, this can lead to approximately the same Casimir force
over the separation region from 1 to $3\,\mu$m as for a lens
with some imperfection and a plate, both described by the Drude
model. It has been known that experimentally it is hard to
determine the position of the point of closest approach
between a lens and a plate on the lens surface with
sufficient precision. Then it remains uncertain what kind
of surface imperfection (if any) is located near the point
of the closest approach. This leads us to the conclusion that
experiments with large spherical lenses are in fact
unsuitable for resolving the problem of the thermal Casimir
force between real metals.

Results similar in spirit are obtained for an Au-coated lens
of centimeter-size radius of curvature interacting with a
semiconductor or dielectric plate. We calculate the Casimir
force between  a perfectly spherical Au-coated lens and
a dielectric (high-resistivity Si) plate with
the neglect of free charge
carriers (in so doing it makes almost no
difference whether the Drude or the plasma model is used
for the description of Au). We show then that approximately
the same Casimir force over the separation region from 1 to
$3\,\mu$m is obtained for an Au sphere with appropriately
chosen surface imperfections and the following models of
a semiconductor plate: 1) High-resistivity Si with included
dc conductivity; 2) Low-resistivity Si with charge carriers
described by the Drude model; 3) Low-resistivity Si with charge carriers
described by the plasma model. Here, free charge carriers
of the Au sphere are described  by the Drude model
in cases 1) and 2), and by the plasma model in the case 3).
Thus, experiments with large spherical lenses are also not helpful
for resolving the problem of dc conductivity of semiconductor or
dielectric materials in the Lifshitz theory.

The structure of the paper is as follows. In Sec.~II we consider
spherical lenses with surface imperfections of different types
and derive the formulations of the PFA applicable for deformed
spherical surfaces. Demonstration of the influence of surface
inperfections on the magnitude of the Casimir force in the
simplest case of ideal metal bodies is contained in Sec.~III.
Section~IV is devoted to the calculation of the Casimir force
between a real metal plate and a real metal lens with surface
imperfections. In Sec.~V similar results are presented for a
real metal lens interacting with a semiconductor or dielectric
plate. Our conclusions and discussions are contained in Sec.~VI.

\section{Proximity force approximation for spherical lenses with
surface imperfections of different types}

As discussed in Sec.~I, the Casimir force should be more
sensitive than the electrostatic force to surface imperfections
that are invariably present on the mechanically polished and
ground surfaces of any lens of centimeter-size curvature radius.
However, in experiments on measuring the Casimir force in the
lens-plate geometry, comparison between the measurement data and theory
is usually performed by means of the simplified formulation of the
PFA assuming perfect sphericity of the lens surface.\cite{2,3,37}
We demonstrate first how this simplified formulation of the
PFA is obtained from the most general formulation.\cite{38}
Then we apply the general formulation of the PFA to lenses with
surface imperfections of different types.

The most general formulation of the PFA represents the Casimir
force between a lens surface $z=z(x,y)$ and a plate $z=0$
as an integral of the Casimir pressures between pairs of
plane surface elements spaced at separations $z=z(x,y)$:
\begin{equation}
F_{sp}(a,T)=\int_{\Sigma}d\sigma P(z,T).
\label{eq1}
\end{equation}
\noindent
Here, $d\sigma$ is the element of plate area, $\Sigma$ is the
projection of the lens onto the plate, $a$ is the shortest
separation between them,  and $P(z,T)$ is the pressure for two
plane parallel plates at a separation $z=z(x,y)$ at temperature $T$.

We choose the origin of a cylindrical coordinate system on
the plane $z=0$ under the lens center. Then for a perfectly
shaped spherical lens of radius of curvature $R$ the coordinate $z$ of any point
of its surface is given by
\begin{equation}
z=R+a-(R^2-\rho^2)^{1/2}, \quad
\rho^2=x^2+y^2.
\label{eq2}
\end{equation}
\noindent
In this case Eq.~(\ref{eq1}) leads to
\begin{eqnarray}
&&
F_{sp}^{\rm perf}(a,T)=2\pi\int_{0}^{\sqrt{2RD-D^2}}\!\!\!\rho d\rho P(z,T)
\nonumber \\
&&~~~~~~~~~
=
2\pi\int_{a}^{D+a}(R+a-z)P(z,T)dz.
\label{eq3}
\end{eqnarray}
\noindent
Keeping in mind that the Casimir pressure is expressed as
\begin{equation}
P(z,T)=-\frac{\partial{\cal F}_{pp}(z,T)}{\partial z},
\label{eq4}
\end{equation}
\noindent
where ${\cal F}_{pp}(z,T)$ is the free energy per unit area of parallel
plates, and integrating by parts in Eq.~(\ref{eq3}), one arrives
at
\begin{eqnarray}
&&
F_{sp}^{\rm perf}(a,T)=2\pi R{\cal F}_{pp}(a,T)
\nonumber \\
&&~~~
-2\pi(R-D){\cal F}_{pp}(D+a,T)
-
2\pi\int_{a}^{D+a}\!\!{\cal F}_{pp}(z,T)dz.
\label{eq5}
\end{eqnarray}
\noindent
We consider centimeter-size spherical lenses satisfying
conditions $a\ll D$, $a\ll R$. For such lenses
${\cal F}_{pp}(D+a,T)\ll{\cal F}_{pp}(a,T)$.
Because of this, one can neglect the second term on the right-hand
side of Eq.~(\ref{eq5}) in comparison with the first.\cite{37}
It can be also shown \cite{37,39} that the first term on the right-hand
side of Eq.~(\ref{eq5}) is larger than the third by a factor of $R/a$.
This allows one to neglect the third term and arrive to what is
called the simplified formulation \cite{37,39} of the PFA
\begin{equation}
F_{sp}^{\rm perf}(a,T)\approx 2\pi R{\cal F}_{pp}(a,T)
\label{eq6}
\end{equation}
\noindent
widely used for both spherical lenses and for spheres [note that for
a semisphere the second term on the right-hand
side of Eq.~(\ref{eq5}) is exactly equal to zero].

The above derivation shows that the PFA in the form of Eq.~(\ref{eq6})
is applicable only at $a/R\ll 1$. For the real metal sphere
(spherical lens) above real metal plate the analytic expressions
for the Casimir force in terms of scattering amplitudes are
available,\cite{37a,37b,37c} but due to computational difficulties
numerical results were obtained only under the condition\cite{37a,37b}
$a/R\geq 0.1$ and under the condition\cite{37c} $a/R\geq 0.053$.
Computations were performed for metals described by simple plasma
and Drude models\cite{37a,37b} and by the generalized plasma and
Drude models taking into account interband transitions of core
electrons.\cite{37c} The relative deviations between the obtained
exact results for the Casimir force and the approximate results
calculated using the PFA in Eq.~(\ref{eq6}) were found to be less
than $a/R$. It was demonstrated\cite{39} also that the PFA
results approach the respective exact results with decreasing
$a/R$. Keeping in mind that for the experiments performed to date
with small spheres $a/R\approx 10^{-3}=0.1$\% and for experiments
with large spherical lenses $a/R\approx 10^{-5}=0.001$\%,
the use of the PFA for the comparison between experiment and
theory is well justified.

Now we consider real lenses with centimeter-size radii of curvature.
Surfaces of such lenses are far from perfect, even excluding the
rms roughness of a few nanometers from consideration.
In optical technology quality of lens surfaces is
characterized \cite{29b,29c,29d,29e,29f,29g,40,41} in
terms of scratch and dig optical surface specification data.
In particular, depending on the quality of the lens used, digs
(i.e., bubbles and pits) with a diameter varying \cite{29b,40}
from $30\,\mu$m to 1.2\,mm are allowed on the surface.
There may also be scratches on the surface with a width
varying \cite{29b,40} from 3 to $120\,\mu$m. The problem of
bubbles on the centimeter-size lens surface should not be
reduced to the fact that lens  curvature radius  $R$ is
determined with some error. The thickness of each bubble or pit should
of course be less than the absolute error in the measurement of lens
radius of curvature (for a lens \cite{31} with $R=15.10\,$cm,
for instance, $\Delta R=0.05\,$cm). The crucial point is that
curvature radii of bubbles and pits can be significantly different,
as compared to $R$.
Surface imperfections with these local radii of curvature, as we
show below, can give a major contribution to the Casimir force.

As the first example we consider the lens with the curvature
radius $R=15\,$cm having a bubble of the radius of curvature
$R_1=25\,$cm and thickness $D_1=0.5\,\mu$m
near the point of the closest approach to the
plate [see Fig.~1(a)].
The radius of the bubble is
determined from $r^2=2R_1D_1-D_1^2\approx 0.25\,\mbox{mm}^2$,
leading to $2r=1\,\mbox{mm}<1.2\,$mm, i.e., less than
a maximum value allowed \cite{40} by the optical surface
specification data.
The respective quantity $d$ defined in Fig.~1(a) is equal to
$d\approx r^2/(2R)\approx 0.83\,\mu$m. Then the flattening
of a lens surface at the point of closest approach to the
plate is $d-D_1\approx 0.33\,\mu$m which is much less than
$\Delta R$.

The general formulation of the PFA (\ref{eq1}) should be applied
taking into account that the surface of the bubble is described
by the equation
\begin{equation}
z=R_1+a-(R_1^2-\rho^2)^{1/2},
\label{eq7}
\end{equation}
\noindent
where $a$ is the distance between the bottom point of the bubble
and the plate [see Fig.~1(a)]. In this notation the surface of the
lens is described by the equation
\begin{equation}
z=R+D_1-d+a-(R^2-\rho^2)^{1/2}.
\label{eq8}
\end{equation}
\noindent
Using Eqs.~(\ref{eq7}) and (\ref{eq8}) one arrives, instead
of Eq.~(\ref{eq3}), at
\begin{eqnarray}
F_{sp}(a,T)&=&2\pi\int_{a+D_1}^{a+D_1-d+D}\!\!\!
(R-z+D_1-d+a)P(z,T)dz
\nonumber \\
&&+
2\pi\int_{a}^{a+D_1}(R_1-z+a)P(z,T)dz.
\label{eq9}
\end{eqnarray}
\noindent
Now we take into consideration that the quantities
$a$, $d$, and $D_1$ are smaller than the error in the
determination of large radii $R$ and $R_1$.
Then one can rearrange Eq.~(\ref{eq9}) to the form
\begin{eqnarray}
&&
F_{sp}(a,T)\approx 2\pi\int_{a+D_1}^{a+D_1-d+D}\!\!\!
(R-z)P(z,T)dz
\nonumber \\
&&~~~~~~
+2\pi R_1\int_{a}^{a+D_1}P(z,T)dz.
\label{eq10}
\end{eqnarray}
\noindent
Here, the first integral on the right-hand side is calculated
similar to Eqs.~(\ref{eq3}) and (\ref{eq5}) leading to
$2\pi R{\cal F}_{pp}(a+D_1,T)$. Calculating the second integral
with the help of Eq.~(\ref{eq4}), one finally obtains
\begin{equation}
F_{sp}(a,T)\approx 2\pi (R-R_1){\cal F}_{pp}(a+D_1,T)+
2\pi R_1{\cal F}_{pp}(a,T).
\label{eq11}
\end{equation}

Now we consider two more examples of surface imperfection,
specifically, a bubble with the curvature radius $R_1=5\,$cm
[see Fig.~1(b)] and a pit with the curvature radius $R_1=12\,$cm
[see Fig.~1(c)]. In both cases the curvature radius of the
lens remains the same $R=15\,$cm. For the bubble we choose
$D_1=1\,\mu$m which results in $r\approx 0.32\,$mm,
$d\approx 0.33\,\mu$m, and $D_1-d\approx 0.67\,\mu$m
in agreement with allowed values.
Equation (\ref{eq11}) is evidently preserved with the new values
of parameters.

Now we deal with the pit shown in Fig.~1(c). Here, the lens surface
near the point of closest approach to the plate is concave up,
i.e., in the direction of the lens center. The related parameters are
$D_1=1\,\mu$m,  $r\approx 0.49\,$mm,
$d\approx 0.8\,\mu$m, and $d+D_1\approx 1.8\,\mu$m.
The pit surface is described by the equation
\begin{equation}
z=a+D_1-R_1+(R_1^2-\rho^2)^{1/2}.
\label{eq12}
\end{equation}
\noindent
Here, $a$ is the separation distance between the plate and the points
of the circle on the lens surface closest to it.
The surface of the
lens is described as
\begin{equation}
z=R+a-d-(R^2-\rho^2)^{1/2}.
\label{eq13}
\end{equation}
\noindent
Repeating calculations that have led to Eq.~(\ref{eq11})
with the help of Eqs.~(\ref{eq12}) and (\ref{eq13}),
we obtain
\begin{equation}
F_{sp}(a,T)\approx 2\pi (R-R_1){\cal F}_{pp}(a,T)+
2\pi R_1{\cal F}_{pp}(a+D_1,T).
\label{eq14}
\end{equation}

It is evident that Eqs.~(\ref{eq11}) and (\ref{eq14}) lead to
significantly different results than the simplified formulation
of the PFA in Eq.~(\ref{eq6}). The reason is that at separations
$a\gtrsim1\,\mu$m we get $D_1\lesssim a$ and all three
contributions on the right-hand side of Eqs.~(\ref{eq11}) and (\ref{eq14})
are of the same order of magnitude. This is confirmed by the results
of numerical computations for both ideal metals (Sec.~III) and for
real materials (Secs.~IV and V).

\section{Demonstration of the role of
surface imperfections for ideal metal bodies}

We begin with the case of an ideal metal lens with surface imperfections
shown in Fig.~1(a,b,c) near the point of closest approach to an
ideal metal plate. The case of ideal metal bodies, although it
disregards real material properties, allows the demonstration of
the entirely geometrical effect of surface imperfections on the
Casimir force.

To perform numerical computations by Eqs.~(\ref{eq11}) and (\ref{eq14})
one needs convenient representation for the Casimir free energy
per unit area, ${\cal F}_{pp}(z,T)$, in the configuration of two
parallel ideal metal plates. The standard expression for this
quantity is given by \cite{2,42}
\begin{equation}
{\cal F}_{pp}(z,T)=\frac{k_BT}{\pi}\sum_{l=0}^{\infty}
{\vphantom{\sum}}^{\!\prime}\int_{0}^{\infty}k_{\bot}dk_{\bot}
\ln(1-e^{-2zq_l}).
\label{eq15}
\end{equation}
\noindent
Here, $k_B$ is the Boltzmann constant, $k_{\bot}$ is the magnitude
of the projection of the wave vector on the plates,
$q_l^2=k_{\bot}^2+\xi_l^2/c^2$, $\xi_l=2\pi k_BTl/\hbar$ with
$l=0,\,1,\,2,\,\ldots$ are the Matsubara frequencies, and the
primed summation sign means that the term with $l=0$ is
multiplied by 1/2.
The most frequently used form of Eq.~(\ref{eq15}) separates it
into the contribution of zero temperature and thermal
correction. For our purpose, however, it is more convenient
to present Eq.~(\ref{eq15}) as a sum of the high-temperature
contribution to the free energy and the correction to it.
For this purpose
we rewrite Eq.~(\ref{eq15}) in terms of a dimensionless
integration variable $y=2aq_l$ and expand the logarithm in a power
series
\begin{equation}
{\cal F}_{pp}(z,T)=-\frac{k_BT}{4\pi z^2}\sum_{l=0}^{\infty}
{\vphantom{\sum}}^{\!\prime}\int_{\tau_z l}^{\infty}ydy
\sum_{n=1}^{\infty}\frac{e^{-ny}}{n}.
\label{eq16}
\end{equation}
\noindent
Here, the dimensionless parameter $\tau_z$ is defined as
$\tau_z=4\pi zk_BT/(\hbar c)$. After performing integration
and then the summation with respect to $l$, the following
result is obtained:
\begin{eqnarray}
&&
{\cal F}_{pp}(z,T)=-\frac{k_BT}{4\pi z^2}\left[
\vphantom{\sum_{n=1}^{\infty}}
\frac{\zeta(3)}{2}\right.
\label{eq17}\\
&&~~~~~~~~~
+\left.
\sum_{n=1}^{\infty}\frac{e^{-\tau_zn}}{n^2(1-e^{-\tau_zn})}
\left(\frac{1}{n}+\frac{\tau_z}{1-e^{-\tau_zn}}\right)\right],
\nonumber
\end{eqnarray}
\noindent
where $\zeta(x)$ is the Riemann zeta function. Note that the
first contribution on the right-hand side of Eq.~(\ref{eq17})
is just the high temperature limit of the free energy.
This is seen if we take into account that
$\tau_z=2\pi T/T_{\rm eff}$, where the effective temperature
is defined from $k_BT_{\rm eff}=\hbar c/(2z)$, and, thus,
$\tau_z\to\infty$ when $T\gg T_{\rm eff}$.

Now we are in a position to compute the Casimir force between the
spherical lens of large curvature radius with bubbles and pits
of different types and a plane plate, both made of ideal metal.
In the literature it is common to use the
simplified formulation of the PFA (\ref{eq6}) in the sphere-plate
geometry for both small spheres of about $100\,\mu$m radii
and large spherical lenses.
\cite{2,17,19,9,26,3,10,11,12,18,20,21,22,23,24,25,27,28,29,30,31,29a,32,33,34}
In doing so the role of bubbles and
pits on the surface of lenses of centimeter-size curvature
radii is neglected. Equation (\ref{eq6}), however,
is not applicable for lenses with large radius of curvature
because it assumes perfect spherical surface.
For such lenses one should use more complicated results like
those in  Eqs.~(\ref{eq11}) and (\ref{eq14}).
To illustrate this fact, we perform
calculations for three typical model imperfections on the
spherical surface near the point of closest approach to the
plate shown in Fig.~1(a,b,c).

We begin with the surface imperfection shown in Fig.~1(a), where the
bottom of the spherical lens is flattened for approximately
$0.33\,\mu$m (see Sec.~II). Computations of the Casimir force
$F_{sp}(a,T)$ in  Eq.~(\ref{eq11}), taking the bubble into
account, and the force
$F_{sp}^{\rm perf}(a,T)$ in  Eq.~(\ref{eq6}) for a lens with
perfectly spherical surface were performed at $T=300\,$K with
the help of Eq.~(\ref{eq17}) over the separation region from
1 to $3\,\mu$m.
Computations at larger separations would be not warranted
because the experimental error quickly increases with the
increase of $a$. Thus, in the most well known measurement\cite{30}
of the Casimir force by using large lens the relative error
at $a=3\,\mu$m was shown\cite{40a} to be larger than 47\%.
In the experiment\cite{31} with Ge test bodies the relative
error in the measured Casimir force exceeds 100\% at
$a\geq 2\,\mu$m due to the error in subtracted residual
electrostatic forces.
The computational results for the ratio
$F_{sp}(a,T)/F_{sp}^{\rm perf}(a,T)$ as a function of separation
are presented in Fig.~2 by the line labeled 1.
As can be seen in Fig.~2, in the presence of the bubble
shown in Fig.~1(a), the use of
Eq.~(\ref{eq6}) for perfect spherical surface instead of
Eq.~(\ref{eq11}) considerably underestimates the magnitude of
the Casimir force. Thus, at separations $a=1.0$, 1.5, 2.0,
2.5, and $3.0\,\mu$m the quantity $F_{sp}/F_{sp}^{\rm perf}$
is equal to 1.458, 1.361, 1.287, 1.233, and 1.193,
respectively, i.e., the underestimation varies from 46\% at
$a=1\,\mu$m to 19\% at $a=3\,\mu$m.

We continue with surface imperfection shown in Fig.~1(b),
where the thickness of an extra bulge
on the spherical lens around its bottom point is
approximately equal to $0.67\,\mu$m (see Sec.~II).
Computations were performed with Eqs.~(\ref{eq6}) and (\ref{eq11})
using Eq.~(\ref{eq17}).
The computed values of the quantity
$F_{sp}(a,T)/F_{sp}^{\rm perf}(a,T)$ as a function of separation
are shown by the line labeled 2 in Fig.~2. It can be seen that
in this case the assumption of perfect sphericity of a lens
surface considerably overestimates the magnitude of the
Casimir force. Thus, at separations $a=1.0$, 1.5, 2.0,
2.5, and $3.0\,\mu$m the values of the
quantity $F_{sp}/F_{sp}^{\rm perf}$
are equal to 0.429, 0.507, 0.580, 0.641, and 0.689,
respectively, i.e., overestimation varies from 57\% at
$a=1\,\mu$m to 36\% at $a=3\,\mu$m.

Finally we consider the surface imperfection in the form of a pit
presented in Fig.~1(c). Here, the deformation of the lens surface
is characterized by the parameter $d+D_1\approx 1.8\,\mu$m.
The computational results using Eqs.~(\ref{eq6}), (\ref{eq14})
and (\ref{eq17}) are shown by the line labeled 3 in Fig.~2.
Once again, the assumption of perfect lens sphericity
significantly overestimates the magnitude of the
Casimir force. Thus, at separations $a=1.0$, 1.5, 2.0,
2.5, and $3.0\,\mu$m the
ratio $F_{sp}/F_{sp}^{\rm perf}$
is equal to 0.314, 0.409, 0.496, 0.570, and 0.627,
respectively, i.e., overestimation varies from 69\% at
$a=1\,\mu$m to 37\% at $a=3\,\mu$m.

Thus, for an ideal metal lens above an ideal metal plate
the use of the PFA in its simplest form (\ref{eq6})
can lead to the Casimir force,
either underestimated or overestimated by many
tens of percent, depending on the character of imperfection
on the lens surface near the point of closest approach to
the plate.
Below we show that for a lens with a centimeter-size
radius of  curvature
and a plate made of real materials the role of imperfections
of the lens surface increases in importance.

\section{Test bodies made of real metal}

At separations above $1\,\mu$m the characteristic frequencies
giving major contribution to the Casimir force are sufficiently
small. Because of this, one can neglect the contribution of
interband transitions and describe the metal of the test bodies
by means of simple Drude model. This leads to the dielectric
permittivity depending on the frequency
\begin{equation}
\varepsilon_{D}(\omega)=1-\frac{\omega_p^2}{\omega(\omega+i\gamma)},
\label{eq18}
\end{equation}
\noindent
where $\omega_p$ is the plasma frequency and $\gamma$ is the
relaxation parameter.

The dielectric permittivity (\ref{eq18}) takes into account
relaxation properties of free electrons by means of the
temperature-dependent relaxation parameter
$\gamma=\gamma(T)$. It is common knowledge that in the local
approximation it correctly describes the interaction of
a metal with the real (classical) electromagnetic field, specifically, in the
quasistatic limit \cite{43} where $\omega\to 0$.
The behavior of $\varepsilon$ as the reciprocal of the frequency
in this limit is the direct consequence\cite{41a} of the classical
Maxwell equations.
It can be said that the Drude dielectric permittivity (\ref{eq18})
is fully justified on the basis of fundamental physical theory and
confirmed in numerous technical applications.
The Drude model (\ref{eq18}) also provides the
correct description of out of thermal
equilibrium  physical phenomena
determined by the fluctuating electromagnetic field
such as radiative heat transfer and near-field
friction.\cite{41b} Because of this, a disagreement of Eq.~(\ref{eq18})
with the experimental data
would be a problem of serious concern.
However, as was noticed in Sec.~I, precise experiments on
measuring the Casimir pressure at separations below $1\,\mu$m
 by means of the micromechanical
torsional oscillator \cite{9,10,11} exclude large thermal
effect in the Casimir force caused by the relaxation
properties of charge carriers in metals. The results of these
experiments are consistent with the plasma model
\begin{equation}
\varepsilon_{p}(\omega)=1-\frac{\omega_p^2}{\omega^2},
\label{eq19}
\end{equation}
\noindent
obtained from Eq.~(\ref{eq18}) by setting $\gamma=0$.

In classical electrodynamics\cite{43,41a} the plasma model
is considered as an approximation valid in the region of
sufficiently high infrared frequencies, where the electric
current is pure imaginary and the relaxation properties do
not play any role. In real (classical) electromagnetic
fields the dielectric permittivity (\ref{eq19}) does not
describe the reaction of a metal on the field in the
limit of quasistatic frequencies.
As was noted above, Maxwell equations lead to
$\varepsilon\sim 1/\omega$ in the limiting case
$\omega\to 0$.
The contradiction between the Lifshitz theory combined
with the Drude model and the experimental data
was widely discussed in the literature,
\cite{2,3,44,45,46,47,48,49,50,51,52} but a
resolution has not yet been achieved. It was also
suggested \cite{53} that there might be some differences
in the reaction of a physical system in thermal equilibrium
with an environment to real fields with nonzero mean value and
fluctuating fields whose mean value is equal to zero. Because of this,
the possibility of measuring the thermal Casimir force
at separations of a few micrometers, where the
predicted results from using Eqs.~(\ref{eq18}) and
(\ref{eq19}) differ up to a factor of two, is of crucial
importance.

We first consider surface inperfections introduced in
Fig.~1(a,b,c) in Sec.~II and compute their impact on the
Casimir force between a lens and a plate, both described
either by the Drude or by the plasma model. In so doing
Eq.~(\ref{eq11}) remains valid for the imperfections of
Fig.~1(a,b) and Eq.~(\ref{eq14}) for the imperfection of
Fig.~1(c). As to the free energy per unit area of two
parallel plates, one should use the following Lifshitz
formula \cite{2,3,4,5} instead of Eq.~(\ref{eq17}):
\begin{equation}
{\cal F}_{pp}(a,T)=\frac{k_BT}{2\pi}
\sum_{l=0}^{\infty}{\vphantom{\sum}}^{\!\prime}
\int_{0}^{\infty}k_{\bot}dk_{\bot}\sum_{\alpha}
\ln(1-r_{\alpha}^2e^{-2aq_l}).
\label{eq20}
\end{equation}
\noindent
Here, $\alpha={\rm TM}$ or TE for the electromagnetic waves with
transverse magnetic and transverse electric polarizations,
respectively, and the reflection coefficients at the
imaginary Matsubara frequencies are given by
\begin{eqnarray}
&&
r_{\rm TM}=r_{\rm TM}(i\xi_l,k_{\bot})=
\frac{\varepsilon(i\xi_l)q_l-k_l}{\varepsilon(i\xi_l)q_l+k_l},
\nonumber \\
&&
r_{\rm TE}=r_{\rm TE}(i\xi_l,k_{\bot})=
\frac{q_l-k_l}{q_l+k_l},
\label{eq21}
\end{eqnarray}
\noindent
where
\begin{equation}
k_l=k(i\xi_l,k_{\bot})=\left[k_{\bot}^2+\varepsilon(i\xi_l)
\frac{\xi_l^2}{c^2}\right]^{1/2}.
\label{eq22}
\end{equation}

For convenience in numerical computations we rearrange Eq.~(\ref{eq20})
in terms of dimensionless wave vector variable $y$ introduced in
Sec.~III and dimensionless Matsubara frequencies
$\zeta_l=\xi_l/\omega_c=\tau_al$, where
$\omega_c=c/(2a)$ is the characteristic frequency:
\begin{equation}
{\cal F}_{pp}(a,T)=\frac{k_BT}{8\pi a^2}
\sum_{l=0}^{\infty}{\vphantom{\sum}}^{\!\prime}
\int_{\zeta_l}^{\infty}ydy\sum_{\alpha}
\ln(1-r_{\alpha}^2e^{-y}).
\label{eq23}
\end{equation}
\noindent
The reflection coefficients are expressed in terms of new variables
in the following way
\begin{eqnarray}
&&
r_{\rm TM}=r_{\rm TM}(i\zeta_l,y)=
\frac{\varepsilon_ly-\sqrt{y^2+\zeta_l^2(\varepsilon_l-1)}}{\varepsilon_ly+
\sqrt{y^2+\zeta_l^2(\varepsilon_l-1)}},
\nonumber \\
&&
r_{\rm TE}=r_{\rm TE}(i\zeta_l,y)=
\frac{y-\sqrt{y^2+\zeta_l^2(\varepsilon_l-1)}}{y+
\sqrt{y^2+\zeta_l^2(\varepsilon_l-1)}},
\label{eq24}
\end{eqnarray}
\noindent
where $\varepsilon_l\equiv\varepsilon(i\omega_c\zeta_l)$.
When the Drude model (\ref{eq18}) is used in computations
we have
\begin{equation}
\varepsilon_l=\varepsilon_l^{D}=1+
\frac{\tilde{\omega}_p^2}{\zeta_l(\zeta_l+\tilde{\gamma})}.
\label{eq25}
\end{equation}
\noindent
Here, the dimensionless plasma frequency and relaxation parameter
are defined as $\tilde{\omega}_p=\omega_p/\omega_c$ and
$\tilde{\gamma}=\gamma/\omega_c$.
In this case the calculated free energy is marked with a
subscript $D$. For the plasma model (\ref{eq19})
\begin{equation}
\varepsilon_l=\varepsilon_l^{p}=1+
\frac{\tilde{\omega}_p^2}{\zeta_l^2},
\label{eq26}
\end{equation}
\noindent
and the Casimir free energy
${\cal F}(a,T)={\cal F}_p(a,T)$.

Now we perform computations of the Casimir force between a real
metal (Au) lens with a surface imperfection around the point of
closest approach to a real metal (Au) plate normalized for the
same force with a perfectly spherical lens.
Note that in real experiments the lens and the plate are usually
made of different materials coated with a metal layer.
For lenses of centimeter-size curvature radius the thickness of
an Au coating can be equal\cite{30} to about $0.5\,\mu$m.
It was shown,\cite{51a} however, that for Au layers of more than
30\,nm thickness the Casimir force is the same as for test bodies
made of bulk Au.
First, we describe the metal
of the lens and the plate by the Drude model with
$\omega_p=9.0\,$eV and $\gamma=0.035\,$eV. Computations are
performed by Eqs.~(\ref{eq11}) and (\ref{eq14}) for
imperfections in Fig.~1(a,b) and 1(c), respectively, with all
parameters indicated in Sec.~II, using
Eqs.~(\ref{eq23})--(\ref{eq25}).
The computational results for the quantity
$F_{sp,D}(a,T)/F_{sp,D}^{\rm perf}(a,T)$
as a function of separation at $T=300\,$K are shown by lines
1, 2, and 3 in Fig.~3 for the surface imperfections presented
in Fig.~1(a), (b), and (c), respectively. These lines are in
qualitative agreement with respective lines in Fig.~2 for
the ideal metal case. Thus, for the surface imperfection
shown in Fig.~1(a) the assumption of a perfectly spherical
surface of the lens leads to an underestimated Casimir
force. As an example, for the imperfection in Fig.~1(a)
the quantity $F_{sp,D}/F_{sp,D}^{\rm perf}$ at
separations of 1 and $3\,\mu$m is equal to 1.176 and
1.097, respectively. Thus, the underestimation of the
Casimir force varies from approximately 18\% to 10\%.
The same quantity at the same respective separations is
equal to 0.4125 and 0.6752 [for the surface imperfection
in Fig.~1(b)] and 0.2951 and 0.6103 [for the surface imperfection
in Fig.~1(c)]. This means that for the imperfection in
Fig.~1(b) the assumption of perfect sphericity leads to an
overestimation of the Casimir force which varies from 59\%
at $a=1\,\mu$m to 32\% at $a=3\,\mu$m. For the surface
imperfection in Fig.~1(c)  the
overestimation  varies from 70\%
 to 39\% when separation increases from 1 to $3\,\mu$m.
 Thus, for real metals described by the Drude model surface
 imperfections of the lens surface play qualitatively the
 same role as for ideal metal lenses. As can be seen in Figs.~2
and 3, the lines labeled 1 for ideal metals and for Drude
metals are markedly different, whereas the respective lines
labeled 2 and 3 in both figures look rather similar.
It is explained by the fact that for the surface imperfection
shown in Fig.~1(a) $D_1=0.5\,\mu$m, while for the imperfections
in Fig.~1(b,c) $D_1=1\,\mu$m. As a result, the influence of
the model of the metal used (ideal metal or the Drude metal)
for the lines labeled 2 and 3 is not so pronounced as for
the line labeled 1.
A few computational results for the quantity
$F_{sp,D}^{\rm perf}$ at different separations for a lens
with $R=15\,$cm are presented in column (a) of Table~I.
They are used below in this section.
For comparison purposes in column (b) of Table~I the same
quantity is computed using the tabulated optical data
for a complex index of refraction\cite{Palik1} extrapolated to
low frequencies by means of the Drude model. As can be seen
in Table~I, the Casimir forces in columns (a) and (b) at
each separation are almost coinciding.
This confirms that at $a\geq 1\,\mu$m the role of interband
transitions is negligibly small, as was noted in the
beginning of this section.

 Now we consider the lens and the plate made of metal
 described by the plasma model (\ref{eq26}) and compute the
 quantity $F_{sp,p}(a,T)/F_{sp,p}^{\rm perf}(a,T)$
 using Eqs.~(\ref{eq11}), (\ref{eq14}) and (\ref{eq23}),
 (\ref{eq24}). It turns out that the computational results
 differ only slightly from respective results shown in
 Fig.~3. Because of this, Fig.~3 is in fact relevant to
 a lens and a plate made of a metal described by the plasma
 model as well. To illustrate minor differences arising
 when the plasma model is used, we present the following
 values of the quantity $F_{sp,p}/F_{sp,p}^{\rm perf}$
 for all three types of surface imperfections shown in
 Fig.~1(a,b,c) at separations $a=1$ and $3\,\mu$m,
 respectively:
 1.17 and 1.092 [imperfection of Fig.~1(a)];
 0.4333 and 0.6916 [imperfection of Fig.~1(b)];
  0.3200 and 0.6300 [imperfection of Fig.~1(c)].
  Comparing these values with the above results obtained
  using the Drude model, we find that relative differences
  vary from a fraction of percent to a few percent.
  In column (c) of Table~I we present several computational
  results for the quantity $F_{sp,p}^{\rm perf}$.
  Column (d) of the same table contains similar results
  computed using the generalized plasma-like
  model\cite{2,3,11} taking into account the interband
  transitions of core electrons. The results of columns (c)
and (d) computed at the same separations are almost
coinciding.

  Now we consider the situation when computational results
  for the Casimir force between a perfectly spherical lens
  above a plate, both described by the plasma model, are
  approximately the same as for a lens with some surface
  imperfection above a plate, both described by the
  Drude model (here and below we use the same Drude
  parameters for Au as already indicated in the text).
  In Fig.~4 the Casimir force $F_{sp,p}^{\rm perf}$ between
  a perfectly shaped lens of $R=15\,$cm radius of curvature and
  a plate versus separation is shown as the solid line
[see also column (c) in Table~I].
 It is computed by Eqs.~(\ref{eq6}), (\ref{eq23}),
(\ref{eq24}), and (\ref{eq26}) at $T=300\,$K.
As an alternative, we assume that there is a surface
imperfection on the lens around the point of closest
approach to the plate shown in Fig.~1(a).
For the parameters of this imperfection (bubble) we
choose $R_1=23\,$cm, $D_1=0.75\,\mu$m which leads to
$r\approx 0.59\,$mm and $d\approx 1.16\,\mu$m.
The flattening of the lens in this case is equal to
$d-D_1\approx 0.41\,\mu$m, i.e., much smaller than
the error in the measurement of lens curvature radius.

Computations of the Casimir force $F_{sp,D}$ between
a lens with this imperfection and a plate as a function
of separation are performed by  Eqs.~(\ref{eq11})
and (\ref{eq23})--(\ref{eq25}).
The computational results are shown in Fig.~4 by the dashed line.
At a few separations, these results are presented in
Table~I, column (e).
As can be seen in Fig.~4, the values of the Casimir force
for a perfectly spherical lens described by the plasma
model are rather close to the force values for a lens with
imperfection described by the Drude model.
For example, using columns (c) and (e) in Table~I, one
obtains that
the relative difference between the two
descriptions
\begin{equation}
\delta F_{sp}(a,T)=\frac{F_{sp,p}^{\rm perf}(a,T)-
F_{sp,D}(a,T)}{F_{sp,p}^{\rm perf}(a,T)}
\label{eq27}
\end{equation}
\noindent
varies from --11\% at $a=1\,\mu$m to 34\% at $a=3\,\mu$m.
Keeping in mind that the error of force measurements
quickly increases with the increase of separation, it
appears impossible to make any definite conclusion on
the model of dielectric properties from the extent of
agreement between the experimental data and theory.

Now we consider the opposite situation, i.e., when the
Casimir force $F_{sp,D}^{\rm perf}$ is approximately equal
to $F_{sp,p}$ for a sphere with some imperfection over
the separation region from 1 to $3\,\mu$m.
The Casimir force $F_{sp,D}^{\rm perf}$ between a
perfectly spherical lens of $R=15\,$cm radius of curvature
and a plate, both described by the Drude model (\ref{eq25}),
was computed as a function of separation by Eqs.~(\ref{eq6})
and (\ref{eq23})--(\ref{eq25}).
The computational results are shown in Fig.~5(a) by the dashed
line [see also column (a) in Table~I].
Large deviation between the solid line in Fig.~4 and
the dashed line in Fig.~5(a) reflects the qualitative difference
between the theoretical descriptions of the Casimir force
by means of the plasma and Drude models.

Approximately the same theoretical results, as shown by the
dashed line in Fig.~5(a), can be obtained, however, for a
lens and plate metal described by the plasma model if the lens
surface possesses some specific imperfection near the point
of closest approach to the plate. In Sec.~II we have considered
only the most simple surface imperfections. There may be more
complicated imperfections on the lens surface, specifically,
different combinations of imperfections shown in Fig.~1(a,b,c).
In Fig.~5(b) we show the surface imperfection on the lens
surface with $R=15\,$cm curvature radius consisting of two
bubbles. The first bubble is of $R_1=3\,$cm radius of curvature.
 It is of the same type as that shown in Fig.~1(b).
The second bubble on the bottom of the first is of
$R_2=19\,$cm curvature radius. It is like that in Fig.~1(a).
{}From Fig.~5(b) one obtains $D_1\approx 1.5\,\mu$m,
 $D_2\approx 0.2\,\mu$m, and $r\approx 0.47\,$mm.
 For the increase of lens thickness at the point of closest
 approach to the plate due to the presence of bubbles,
 we find $0.74\,\mu$m which is much smaller than the error in
 the measurement of the lens radius of curvature.
 The Casimir force between the spherical lens with two
 bubbles and a plate is given by the repeated application
 of Eq.~(\ref{eq11}) to each of the bubbles
 \begin{eqnarray}
 &&
 F_{sp}(a,T)=2\pi(R_1-R_2){\cal F}_{pp}(a+D_2,T)+
 2\pi R_2{\cal F}_{pp}(a,T)
 \nonumber \\
 &&~~~~~~~~~~~~
 +2\pi(R-R_1){\cal F}_{pp}(a+D_1,T).
 \label{eq28}
 \end{eqnarray}

 We performed numerical computations of the Casimir force
$F_{sp,p}$ as a function of separation using Eqs.~(\ref{eq23}),
(\ref{eq24}), (\ref{eq26}), and (\ref{eq28}).
The computational results are shown in Fig.~5(a) by the solid
line. At a few separation distances these results are presented
in TableI1, column (f).
As can be seen in Fig.~5(a), the theoretical lines
computed for a perfectly spherical lens using the Drude model
and for a lens with a surface imperfection using the plasma
model are rather close.
Quantitatively from columns (a) and (f) in Table~I
one obtains that the quantity
\begin{equation}
\delta \tilde{F}_{sp}(a,T)=\frac{F_{sp,D}^{\rm perf}(a,T)-
F_{sp,p}(a,T)}{F_{sp,D}^{\rm perf}(a,T)}
\label{eq29}
\end{equation}
\noindent
varies from --10\% at $a=1\,\mu$m to 26\% at $a=3\,\mu$m.
Such small differences do not allow experimental resolution
between alternative theoretical descriptions of the lens and
plate material by means of the Drude and plasma models.
The reason is that in
experiments with lenses of centimeter-size radius of curvature
at large separations, as explained in Sec.~III,
the experimental error exceeding a few tens of
percent is expected.

\section{Metallic or semiconductor lens above a semiconductor plate}

As mentioned in Sec.~I, the account of relaxation properties of free
charge carriers in semiconductor and dielectric materials also
creates problems for the  theoretical description of the thermal
Casimir force. Here, most of experiments\cite{12,21,22,25,54}
were performed with an
Au-coated sphere of about $100\,\mu$m radius above a semiconductor
plate, and only
one \cite{31} with a Ge spherical lens of $R=15.1\,$cm above a
Ge plate. The measurement data of the two experiments \cite{12,13}
are inconsistent with the inclusion of dc conductivity into
a model of the dielectric response for high-resistivity
semiconductors with the concentration of charge carriers below
critical (i.e., for semiconductors of dielectric type whose
conductivity goes to zero when temperature vanishes) and
for dielectrics. The question  of how to describe free charge
carriers of low-resistivity semiconductors in the Lifshitz
theory (e.g., by means of the Drude or plasma model)
also remains unsolved. One may hope that these problems
can be solved in experiments on measuring the Casimir force
between large Au-coated or semiconductor
lenses above semiconductor plates.
Below we show, however, that invariably present imperfections of
lens of large radius of curvature
do not allow one to discriminate between
the different theoretical models.

We start with a perfectly spherical Au-coated lens of
$R=15\,$cm curvature radius above a Si plate.
Within the first model we describe a high-resistivity Si plate
as a true dielectric with the dielectric permittivity
$\varepsilon_{\rm Si}(\omega)$ determined from the
tabulated optical data \cite{55} for Si samples with the
resistivity $\rho_0=1000\,\Omega\,$cm.
In so doing $\varepsilon_{\rm Si}(0)=11.66<\infty$.
This model is an approximation because it disregards the dc
conductivity of Si.
The computational results for the Casimir force between
a lens and a plate computed using Eqs.~(\ref{eq6}),
(\ref{eq23}), and (\ref{eq24}) with
$\varepsilon_l=\varepsilon_{\rm Si}(i\omega_c\zeta_l)$
at $T=300\,$K are shown by the upper solid line in
Fig.~6. These results are almost independent of whether
the Drude or the plasma model is used for the description
of the lens metal. Specifically, the relative difference
in force magnitudes due to the use of the Drude or
plasma models decreases from 0.22\% to 0.031\% when
the separation
distance increases from 1 to $3\,\mu$m.

Within the second model we consider the same high-resistivity
Si plate, but take the dc conductivity into account.
Then the dielectric permittivity can be presented
in the form
\begin{equation}
\varepsilon_{\rm Si}^{dc}(\omega)=\varepsilon_{\rm Si}(\omega)+
i\frac{4\pi\sigma_0}{\omega},
\label{eq30}
\end{equation}
\noindent
where $\sigma_0=\sigma_0(T)$ is the static conductivity.
In the local approximation the permittivity (\ref{eq30})
correctly describes the reaction of semiconductors on
real electromagnetic fields.
In this case computations using Eqs.~(\ref{eq6}),
(\ref{eq23}), and (\ref{eq24}) result in the dotted line  in
Fig.~6. Note that the computational results do not depend on the
value of $\sigma_0$ in Eq.~(\ref{eq30}), but only from the
fact that $\sigma_0\neq 0$.
The dotted line in Fig.~6 is also almost independent on
whether the Drude or the plasma model is used for the
description of a lens metal.

As the third and fourth models we consider Si plate made of
low-resistivity B doped Si with the concentration of
charge carriers above the critical value. \cite{21}
This is a semiconductor of metallic type whose conductivity does
not go to zero when the temperature vanishes. We present the
dielectric permittivity of such a plate in the form
(the third model)
\begin{equation}
\varepsilon_{\rm Si}^{D}(\omega)=\varepsilon_{\rm Si}(\omega)-
\frac{\omega_{p,\,\rm Si}^2}{\omega(\omega+i\gamma_{\rm Si})},
\label{eq31}
\end{equation}
\noindent
where the values of the Drude parameters are \cite{21}
$\omega_{p,\,\rm Si}\approx 0.46\,$eV and
$\gamma_{\rm Si}\approx 0.099\,$eV,
or in the form (the fourth model)
\begin{equation}
\varepsilon_{\rm Si}^{p}(\omega)=\varepsilon_{\rm Si}(\omega)-
\frac{\omega_{p,\,\rm Si}^2}{\omega^2}.
\label{eq32}
\end{equation}
\noindent
The results of the computations  using Eqs.~(\ref{eq6}),
(\ref{eq23}), (\ref{eq24}), and either (\ref{eq31}) or
(\ref{eq32}) are presented in Fig.~6 by the dashed and lower solid
lines, respectively. Similar to models used for the description
of low-resistivity Si, the metallic lens was described by the
Drude model [when Si was described by Eq.~(\ref{eq31})] or by the
plasma  model [for the dielectric permittivity of Si in Eq.~(\ref{eq32})].
Note that the magnitudes of the Casimir forces given by the second
and third models (the dotted and dashed lines in Fig.~6,
respectively) are very close. When separation increases from 1 to
$3\,\mu$m, the relative differences between the dotted and dashed
lines decrease from 4.8\% to 0.75\%, respectively.
Figure 6 demonstrates that the magnitudes of the Casimir force
between an Au lens and a Si plate may vary over a wide range
depending on the choice of a Si sample and theoretical model used.

Now we present a few computational results for the Casimir force
between an Au-coated lens and a Si plate when the  Si is described by
the different models listed above and the lens may have surface imperfections
around the point of closest approach to the plate.
We first consider the plate made of dielectric Si (the first model)
described by the dielectric permittivity
$\varepsilon_{\rm Si}(\omega)$ and the Au-coated lens of perfect
sphericity. The Casimir force in this case is shown by the solid
line in Fig.~7(a) which was already presented in Fig.~6 as the
upper solid line. Now let the plate be made of low-resistivity
Si described by the Drude dielectric permittivity (\ref{eq31})
(the third model) and the lens possess the surface imperfection
shown in Fig.~1(b) with the parameters $R_1=12.5\,$cm,
$D_1=1\,\mu$m, $r\approx 0.5\,$mm. The results of the numerical
computations for the Casimir force using Eqs.~(\ref{eq11}),
(\ref{eq23})--(\ref{eq25}), and (\ref{eq31}) are shown by the
dashed line in Fig.~7(a). As is seen from this figure, the dashed
line is almost coinciding with the solid one. Thus, the relative
deviation of the Casimir force for a lens with surface imperfection
from the force with a perfectly spherical lens [defined similar to
 Eqs.~(\ref{eq27}) and (\ref{eq29})] varies from --2\% to 13\%
 when separation increases from 1 to $3\,\mu$m.
 Because of this, with lenses of large radius of curvature it is not
 possible to experimentally resolve between the case of
 high-resistivity (dielectric) Si described by the finite
dielectric permittivity $\varepsilon_{\rm Si}(\omega)$ and
low-resistivity Si described by the Drude model. Almost the same
Casimir forces, as shown by the dashed line in Fig.~7(a),
are obtained for a plate made of high-resistivity Si with dc
conductivity included in accordance with Eq.~(\ref{eq30})
(the second model) if the lens has an imperfection
shown in Fig.~1(b) with the parameters $R_1=13\,$cm,
$D_1=1\,\mu$m, $r\approx 0.5\,$mm.
In this case the Casimir force for a lens with surface imperfection
deviates
from the force for a perfectly spherical lens
by --3\% at $a=1\,\mu$m and 14\% at $a=3\,\mu$m.

The last, fourth, model to discuss is of the plate made of
low-resistivity Si described by the plasma dielectric permittivity
(\ref{eq32}). In this case we consider an Au-coated sphere with
surface imperfection (two bubbles) shown in Fig.~5(b).
The parameters of the imperfection are the following:
$R_1=1.5\,$cm, $R_2=21\,$cm, $D_1=2\,\mu$m, $D_2=0.2\,\mu$m
leading to $r\approx 0.28\,$mm. Computations of the Casimir
force are performed using Eqs.~(\ref{eq23}),
(\ref{eq24}), (\ref{eq26}), and (\ref{eq32}).
The computational results are shown as the dashed line in
Fig.~7(b). In the same figure the solid line reproduces the
Casimir force acting between a perfectly spherical lens and
a plate made of dielectric Si. The relative differences between
the dashed and solid lines in Fig.~7(b) vary from --8\% to
23\% when the separation increases from 1 to $3\,\mu$m.
Thus, experimentally it would be not possible to distinguish
between the cases when the lens surface is perfectly spherical
and the plate is made of dielectric Si, and when the lens surface
has an imperfection, but Si plate is of low-resistivity and
is described by the plasma model.

In the end of this section we briefly consider the spherical lens
of $R=15.1\,$cm radius made of intrinsic Ge above the plate made
of the same semiconductor.\cite{31} In this experiment,
Eq.~(\ref{eq6}) was used\cite{31} for the comparison between the
measurement data and theory. As two simple examples we consider
that the Ge lens has a bubble either of the radius of curvature
$R_1=22\,$cm and thickness $D_1=0.09\,\mu$m or
$R_1=10\,$cm and thickness $D_1=0.2\,\mu$m
near the point of closest approach to a Ge plate [see Fig.~1(a)
and (b), respectively]. The radii of the two bubbles are
coinciding and equal to $r=0.2\,$mm leading to the diameter of
each of the bubbles $2r=0.4\,$mm (see Sec.~II). The obtained
value should be compared with limitations imposed by the
scratch/dig optical surface specification data of the
used\cite{31} Ge lens of ISP optics, GE-PX-25-50.
According to the information from the producer,\cite{530a}
this lens has the surface quality 60/40. The latter means that
0.4\,mm is just the maximum diameter of bubbles allowed.
It is also easily seen that the flattening of the lens surface
in Fig.~1(a) or the swelling up in Fig.~1(b) due to bubbles are
much less than the absolute error of $R$ equal to\cite{31}
$\Delta R=0.05\,$cm. Really, with the above parameters
$d\approx 0.13\,\mu$m. As a result, the flattening of the lens surface
in Fig.~1(a) is given by $d-D_1\approx 0.04\,\mu$m and
the swelling up in Fig.~1(b) is given by $D_1-d\approx 0.07\,\mu$m.
In the presence of bubbles the Casimir force should be calculated
not by Eq.~(\ref{eq6}) but  by Eq.~(\ref{eq11}).
Computations using the dielectric permittivity\cite{530b} of
intrinsic Ge show that for the used parameters of the bubble in
Fig.~1(a) Eq.~(\ref{eq11}) leads to larger magnitudes of the
Casimir force by 15\% and 10\% than Eq.~(\ref{eq6}) at separations
$a=0.6$ and $1\,\mu$m, respectively. On the opposite, for the bubble in
Fig.~1(b) the use of Eq.~(\ref{eq11}) instead of Eq.~(\ref{eq6})
results in smaller magnitudes of the Casimir force by 19\% and
14\% at the same respective separations.

\section{Conclusions and discussion}

In the foregoing we have investigated the impact of imperfections,
which are  invariably present on lens surfaces of centimeter-size
 radius of curvature, on the Casimir force in the lens-plate geometry.
We have demonstrated that if an imperfection in the form of a bubble
or a pit is located near the point of the closest approach of a
lens and a plate, the impact on the Casimir force can be
dramatic. We first considered a metal-coated lens above a
metal-coated plate. It was shown that the Casimir force between
a perfectly spherical lens and a plate, both described by the
plasma model, can be made approximately equal to the force
between a sphere with some surface imperfection and a plate,
both described by the Drude model. Similarly,
the Casimir force computed for
a perfectly spherical lens and a plate described by the Drude model
can be approximately equal to the force computed for a lens with
surface imperfection and a plate described by the
plasma model. In both cases the approximate equality of forces
in the limits of the error of force measurements was found
over a wide range of separations from 1 to $3\,\mu$m.
The absolute impact of surface imperfections on the lens
surfaces of centimeter-size radii of curvature on the Casimir
force is on the order of a few tens of percent for both ideal and real
metals. Surface imperfections can lead to both a decrease and
an increase of the force magnitude. These conclusions obtained by
simultaneous consideration of the Drude and plasma models are of
major importance for experiments aiming to discriminate between
the predictions of both approaches at separations above $1\,\mu$m
and to resolve the long-term controversy in the theoretical
description of thermal Casimir forces.

The above conclusions were obtained using the spatially local Drude
and plasma dielectric functions. The possible impact of nonlocal
dielectric permittivity on the thermal Casimir force between
metallic test bodies was investigated in detail in the literature.
Specifically, it was shown\cite{53a} that even for metal coatings
thinner than the mean free path of electrons in the bulk metal,
the relative difference in the thermal Casimir forces computed
using the local Drude model and nonlocal permittivities is less
than a few tenth of a percent ($\approx 0.2$\% at $a=100\,$nm
and decreases with the increase of separation). For thicker metal
coatings used in experiments the contribution of nonlocal effects
to the thermal Casimir force further decreases. This is explained
by the fact that the use of nonlocal dielectric permittivities
leads\cite{53b,53c} to the same equality,
$r_{\rm TE}(0,k_{\bot})=0$, as does the Drude model (\ref{eq18}).
Thus, there is no need to consider nonlocal dielectric functions
in connection with surface imperfections of lenses with
centimeter-size curvature radii. For other fluctuation
phenomenon, as radiative heat transfer, it was also
calculated\cite{41b} that at separations between two metallic
semispaces $a>100\,$nm the contribution of nonlocal effects
into the heat flux is very small.

Similar results were obtained for an Au-coated lens of
centimeter-size radius of curvature above a Si plate.
It was shown that different models for the description of
charge carriers in Si (dielectric Si, high-resistivity Si
with account of dc conductivity, low-resistivity Si described
by the Drude model, and low-resistivity Si described
by the plasma model) lead to different theoretical predictions
for the Casimir force between a perfectly spherical
Au-coated lens and a Si plate. However, by choosing an
appropriate imperfection,
well within the optical surface specification data,
on the surface of the lens at the
point of closest approach to the plate it is possible to
obtain approximately the same Casimir forces in all the
above models over the separation region from 1 to $3\,\mu$m.

The above results show that the presently accepted approach to
the comparison of the data and theory in
experiments\cite{30,31,29a,29aa}
measuring the Casimir force by means of lenses with
centimeter-size radii of curvature might be not
sufficiently justified.
In these experiments the Casimir force is computed using
the simplest formulation of the PFA in Eq.~(\ref{eq6}), i.e.,
under an assumption of perfect sphericity of the lens surface.
According to our results, however, Eq.~(\ref{eq6}) is not
applicable in the presence of surface imperfections which
are invariably present on lens surfaces.
In fact, for reliable comparison between the measurement
data and theory it would be necessary, first, to determine
the position of the point of closest approach to the plate
on the lens surface with a precision of a fraction of
micrometer. Then one could investigate the character of
local imperfections in the vicinity of this point
microscopically and derive an approximate formulation of
the PFA like in Eqs.~(\ref{eq11}), (\ref{eq14}) or
(\ref{eq28}). Thereafter the measurement data could be
compared with theory with some degree of certainty.
It is unlikely, however, that sufficiently precise
determination of the point of closest approach to the
plate is possible for lenses of centimeter-size
radii of curvature.
The possibility to return to the same point of
closest approach in repeated measurements is all the
more problematic. Because of this, one can conclude
that measurements of the Casimir force using lenses of
centimeter-size radii of curvature do not allow an
unambiguous  comparison to
theory, and are not reproducible (see Refs.\cite{30,29a,29aa}
whose results are mutually contradictory).
According to our results, even if the measurement data
for the Casimir force are not consistent with any
theoretical model under an assumption of perfect
sphericity of the lens surface, there might be
different types of surface imperfections leading to
the consistency of the data with several theoretical
approaches.

We emphasize that only a few simple surface imperfections in
the form of bubbles and pits are considered in this paper.
There are many other imperfections of a more complicated
shape (including scratches) which are allowed by the
optical surface specification data and may strongly impact
on the Casimir force between a centimeter-size lens and a plate.
Such imperfections are randomly distributed on lens surfaces
and some of them can be located in the immediate region of
the point of closest approach to the plate. The role of
 Au coatings used in measurements of the Casimir force,
 should be investigated as well in the presence of surface
 imperfections. Metallic coating of about $0.5\,\mu$m
 thickness \cite{30} might lead to a decrease of thicknesses of
 bubbles and depths of pits, but to an increase of their
 diameters. The latter, however, influences the magnitude
 of the Casimir force most dramatically.

The above discussed fundamental problem arising in the
measurements of the Casimir force using lenses of
centimeter-size curvature radii does not arise for microscopic
spheres of about $100\,\mu$m radii used in numerous
experiments by different authors performed with the help of
an atomic force microscope
\cite{2,17,19,3,12,18,20,21,22,23,27,28,29,34,54}
and micromechanical torsional
oscillator.\cite{2,9,26,3,10,11,24,25}
For instance, for microscopic polystyrene spheres
made by the solidification
from the liquid phase the minimization of surface energy
leads to perfectly smooth spherical surfaces
due to surface tension. The surface quality of such
spheres after metallic coating was investigated using
a scanning electron microscope \cite{2,17,3}
and did not reveal any bubbles or scratches.
Spheres of microscopic size have been successfully used
\cite{9,10,11} to exclude large thermal effect in the
Casimir force at separations below $1\,\mu$m.
They are, however, not suitable for measurements of
the thermal effect at large separations of a few
micrometers because the Casimir force is proportional
to the sphere radius and rapidly decreases with the
increase of separation. Keeping in mind the above discussed
fundamental problem arising for spherical
lenses of centimeter-size radius of curvature,
the only remaining candidate for the measurement
of thermal effect in the Casimir force at
micrometer separations is the classical Casimir
configuration of two parallel plates.\cite{58}

%%%%%%%%%%%%%%%%%%%%%%%%%%%%%%%%%%%%%%%%%%%%%%%%%%%%%%%%%%%%%%%%%%%%
\section*{Acknowledgments}

G.L.K.\ and V.M.M.\ are grateful to the Federal University of
Para\'{\i}ba  for kind hospitality.
The work of V.B.,\ G.L.K.\ V.M.M.\ and C.R.\ was supported
by CNPq (Brazil).
G.L.K.\ was also
partially supported by the grant of the Russian Ministry of
Education P--184.
U.M.\ was supported by NSF grant PHY0970161 and DOE grant
DEF010204ER46131.
%%%%%%%%%%%%%%%%%%%%%%%%%%%%%%

%%%%%%%%%%%%%%%%%%%%%%%%%%%%%%%%%%%%%%%%%
%%%%%%%%%%%%___FIGURES__%%%%%%%%%%%%%%%%
%%%%%%%%__FIGURE__1__%%%%%%%%%%%%%%%%%%%%%%%%%%%%%%%%%%%%
\begin{figure*}[h]
\vspace*{-3cm}
\centerline{
\includegraphics{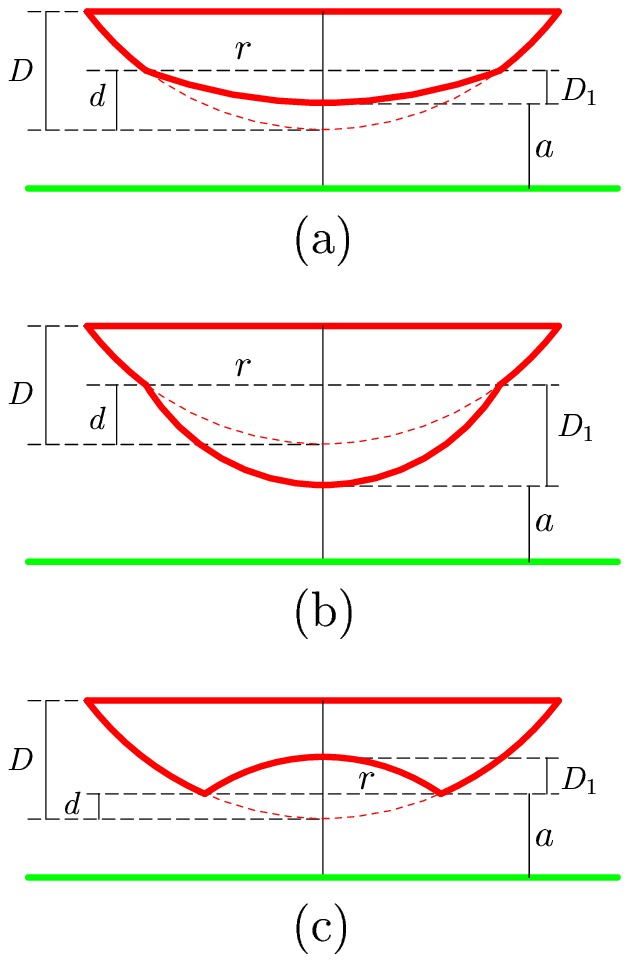}
}
\vspace*{-13cm}
\caption{(Color online)
The configuration of a spherical lens with radius of curvature $R$
possessing a surface imperfection around the point of closest
approach to a plate. (a) The bubble radius of curvature is $R_1>R$.
(b) The bubble radius of curvature is $R_1<R$.
(c) The pit radius of curvature is $R_1<R$.
The relative sizes of the lens and imperfection are not shown
to scale.} \label{fig1}
\end{figure*}
%%%%%%%%%%%%%%%%%%%%%%%%%%%%%%%%%%%%%%%%%%%%%%%%%%%%%%%%%
%%%%%%%%__FIGURE__2__%%%%%%%%%%%%%%%%%%%%%%%%%%%%%%%%%%%%
\begin{figure*}[h]
\vspace*{-10cm}
\centerline{\hspace{2.5cm}\includegraphics{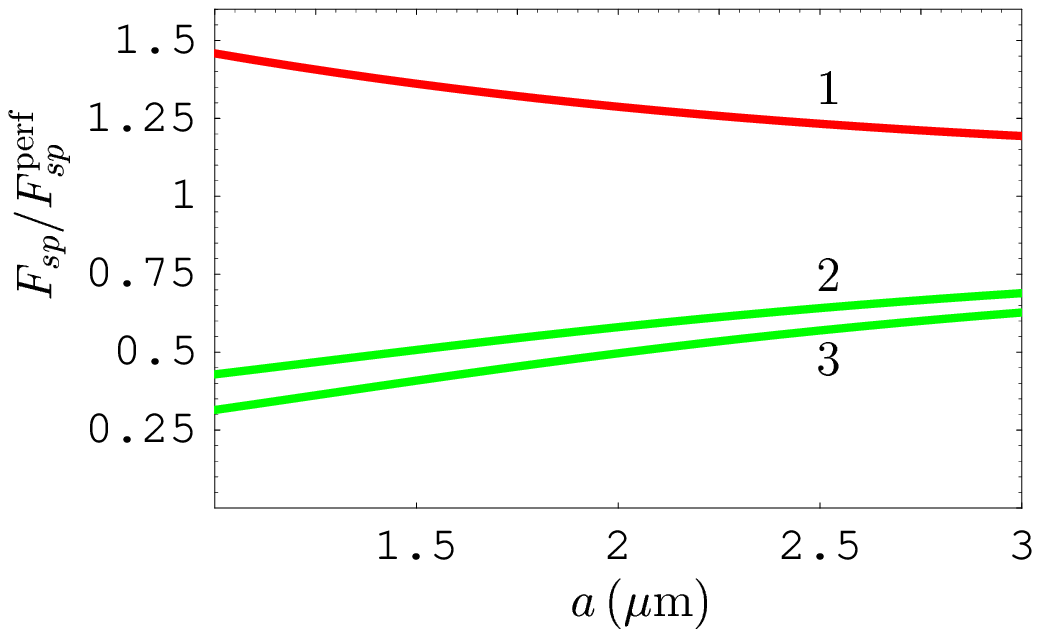}}
\vspace*{-4cm}
 \caption{(Color online)
 The normalized Casimir force acting between an
 ideal metal spherical lens with
surface imperfections of different types and an
ideal metal plate as
a function of separation. Lines 1, 2, and 3 are for the
surface imperfections shown in Fig.~1(a,b,c),
respectively.} \label{fig2}
\end{figure*}
%%%%%%%%%%%%%%%%%%%%%%%%%%%%%%%%%%%%%%%%%%%%%%%%%%%%%%%%%
%%%%%%%%__FIGURE__3__%%%%%%%%%%%%%%%%%%%%%%%%%%%%%%%%%%%%
\begin{figure*}[h]
\vspace*{-10cm}
\centerline{\hspace{2.5cm}\includegraphics{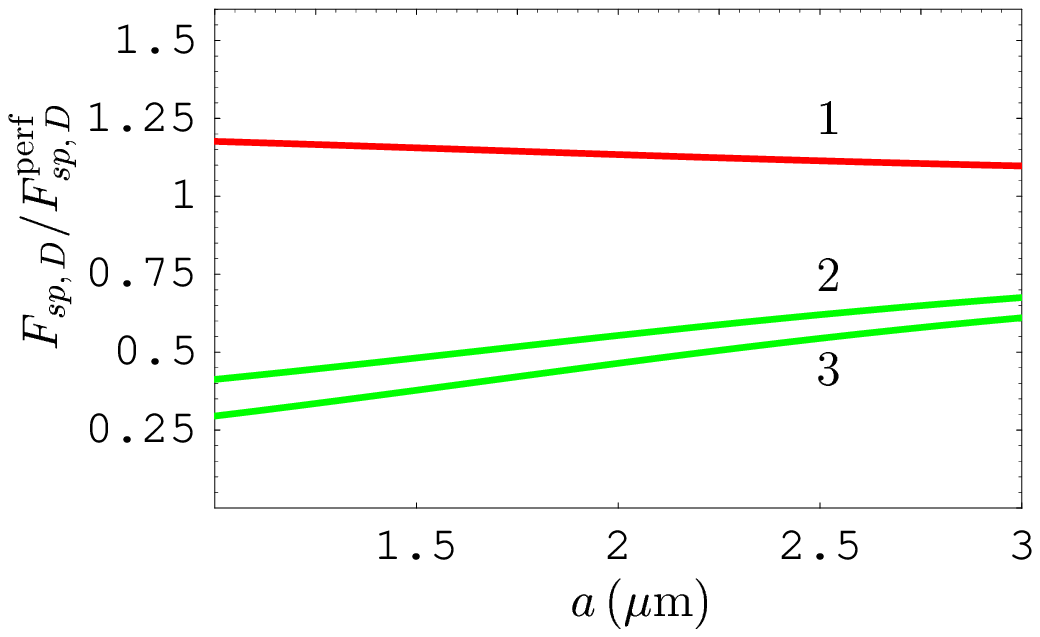}}
\vspace*{-4cm}
 \caption{(Color online)
 The normalized Casimir force acting between an
 Au-coated spherical lens with
surface imperfections of different types and an
Au plate as
a function of separation. Au is described by the Drude
model. Lines 1, 2, and 3 are for the
surface imperfections shown in Fig.~1(a,b,c),
respectively.} \label{fig3}
\end{figure*}
%%%%%%%%%%%%%%%%%%%%%%%%%%%%%%%%%%%%%%%%%%%%%%%%%%%%%%%%%
%%%%%%%%__FIGURE__4__%%%%%%%%%%%%%%%%%%%%%%%%%%%%%%%%%%%%
\begin{figure*}[h]
\vspace*{-10cm}
\centerline{\hspace{2.5cm}\includegraphics{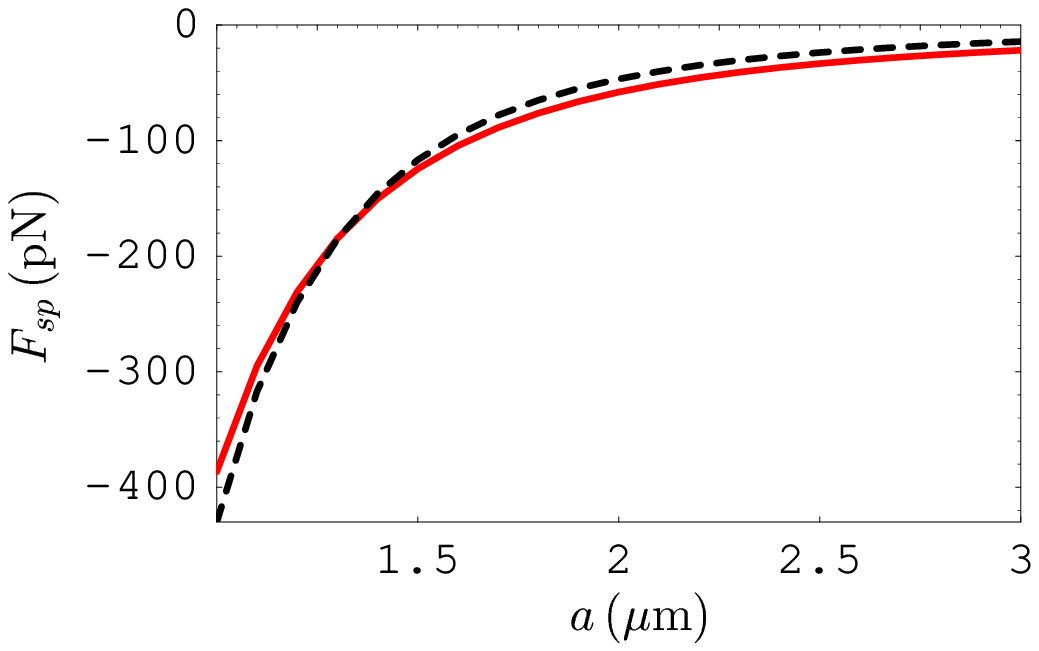}}
\vspace*{-4cm}
 \caption{(Color online)
 The Casimir force between a perfectly spherical lens and a plate,
 both described by the plasma model, versus separation is shown
 by the solid line. The dashed line shows the same force
 between a sphere with some surface imperfection and a plate,
 both described by the Drude model. See text for further discussion.
 } \label{fig4}
\end{figure*}
%%%%%%%%%%%%%%%%%%%%%%%%%%%%%%%%%%%%%%%%%%%%%%%%%%%%%%%%%
%%%%%%%%__FIGURE__5__%%%%%%%%%%%%%%%%%%%%%%%%%%%%%%%%%%%%
\begin{figure*}[h]
\vspace*{-2cm}
\centerline{\hspace{2.cm}\includegraphics{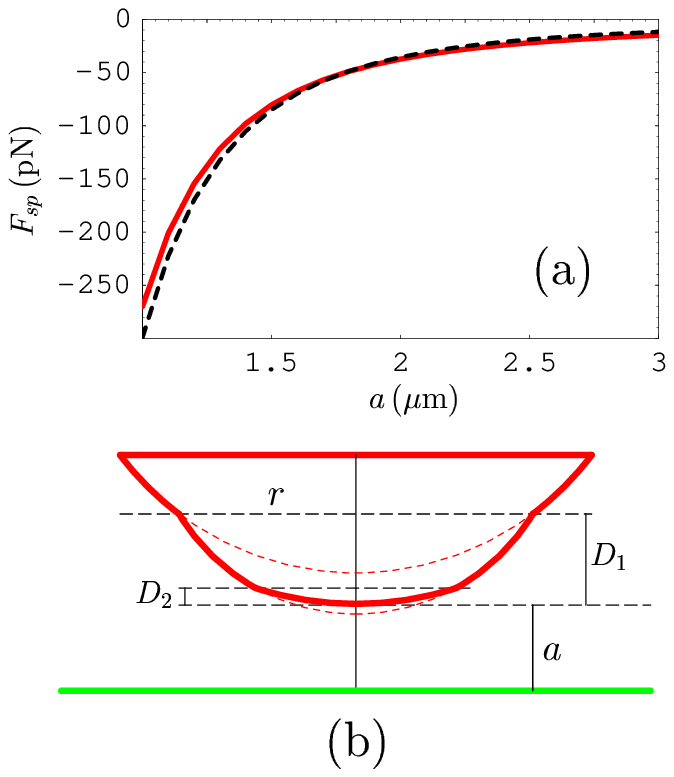}}
\vspace*{-17cm}
 \caption{(Color online)
 (a) The Casimir force between a perfectly spherical lens and a plate,
 both described by the Drude model, versus separation is shown
 by the dashed line. The solid line shows the same force
 between a sphere with some surface imperfection and a plate,
 both described by the plasma model. See text for further discussion.
 (b) The configuration of a spherical lens with radius of curvature $R$
possessing a surface imperfection in the form of two bubbles
around the point of closest approach to a plate.
The relative sizes of the lens and imperfection are shown
not to scale.
 } \label{fig5}
\end{figure*}
%%%%%%%%%%%%%%%%%%%%%%%%%%%%%%%%%%%%%%%%%%%%%%%%%%%%%%%%%
%%%%%%%%__FIGURE__6__%%%%%%%%%%%%%%%%%%%%%%%%%%%%%%%%%%%%
\begin{figure*}[h]
\vspace*{-10cm}
\centerline{\hspace{2.5cm}\includegraphics{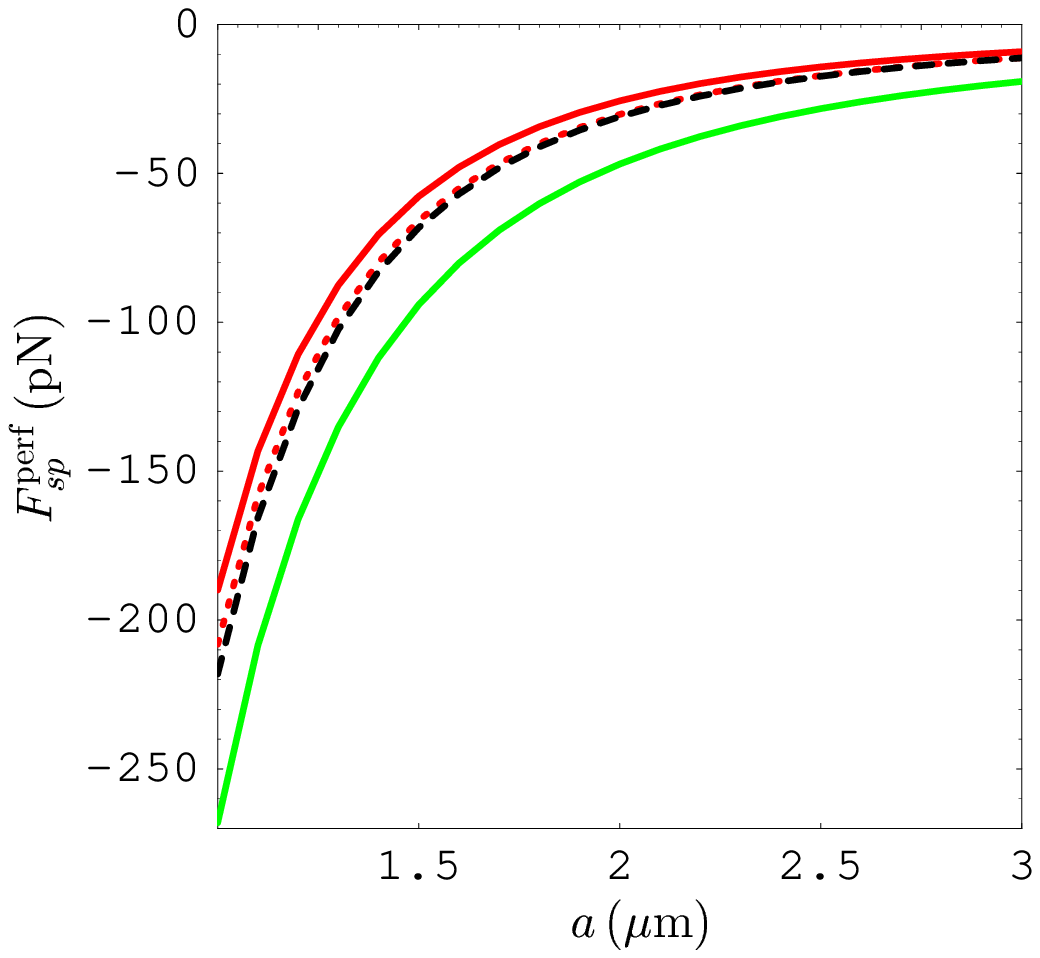}}
\vspace*{-6cm}
 \caption{(Color online)
 The Casimir force between a perfectly spherical
 Au-coated lens and Si plate as a function of separation
 for dielectric Si (the upper solid line), dielectric
 Si with dc conductivity included (the dotted line),
 low-resistivity Si described by the Drude model
 (the dashed line), and  low-resistivity Si described
by the plasma model (the lower solid line).
 } \label{fig6}
\end{figure*}
%%%%%%%%%%%%%%%%%%%%%%%%%%%%%%%%%%%%%%%%%%%%%%%%%%%%%%%%%
%%%%%%%%__FIGURE__7__%%%%%%%%%%%%%%%%%%%%%%%%%%%%%%%%%%%%
\begin{figure*}[h]
\vspace*{-2cm}
\centerline{\hspace{1.5cm}\includegraphics{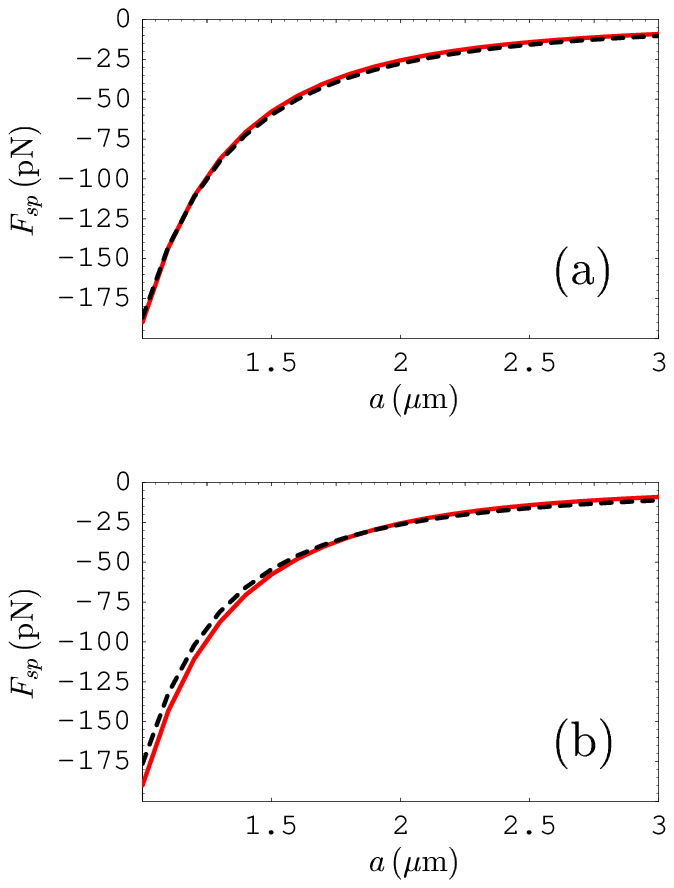}}
\vspace*{-17cm}
 \caption{(Color online)
The Casimir force between a perfectly spherical
Au-coated lens and Si plate made of dielectric Si
versus separation is shown by the solid lines.
The dashed lines show the Casimir force between an
Au-coated lens with some specific surface imperfections
and a plate made of low-resistivity Si where Si is
described (a) by the Drude model and (b) by the plasma
model.
 } \label{fig7}
\end{figure*}
%%%%%%%%%%%%%%%%%%%%%%%%%%%%%%%%%%%%%%%%%%%%%%%%%%%%%%%%%
%\end{document}
%%%%%%%%%%%%__TABLE__%%%%%%%%%%%%%%%%%%%%%%%%%%%%
\begingroup
\squeezetable
\begin{table}
\caption{The values of the Casimir force between an Au-coated
sphere of $R=15\,$cm radius of curvature and an Au-coated
plate computed at $T=300\,$K for (a) Au described by the Drude
model and a perfectly shaped lens;
(b) Au described by the tabulated optical data extrapolated by
means of the Drude model and a perfectly shaped lens;
(c) Au described by the plasma model and a perfectly shaped lens;
(c) Au described by the generalized plasma-like model and a perfectly
shaped lens;
(e) Au described by the Drude model and a lens with surface
imperfection shown in Fig.~1(a);
(e) Au described by the plasma model and a lens with surface
imperfection shown in Fig.~5(b). See text for the parameters
of lens imperfections.
}
\begin{ruledtabular}
\begin{tabular}{ccccccc}
\multicolumn{1}{c}{$a$}
&\multicolumn{6}{c}{$F_{sp}(a)$ (pN)}\\
\cline{2-7}
($\mu$m) & (a) & (b) & (c) & (d) & (e) & (f) \\
\cline{1-7}
1.0 & --299.08& --299.38& --386.56& --386.64& --430.34&--269.93\\
1.5 &--84.914&--84.953&--124.44&--124.44&--116.72&--80.423\\
2.0 &--35.540&--35.548&--57.984&--57.985&--46.681&--37.298\\
2.5 &--18.874&--18.876&--33.330&--33.331&--27.787&--21.961\\
3.0 &--11.744&--11.745&--21.830&--21.830&--14.304&--14.847
 \\
\end{tabular}
\end{ruledtabular}
\end{table}
\endgroup
%%%%%%%%%%%%%%%%%%%%%%%%%%%%%%%%%%%%%%%%%%%%%%%%%%
\end{document}